\documentclass[doublecol]{epl2} 
\usepackage{amsmath}
\usepackage{amsfonts}
\def\tsigma{\widetilde{\sigma}}

\newcommand{\be}{\begin{equation}}
\newcommand{\ee}{\end{equation}}
\newcommand{\ba}{\begin{eqnarray}}
\newcommand{\ea}{\end{eqnarray}}
\newcommand{\bw}{\begin{widetext}}
\newcommand{\ew}{\end{widetext}}

\newcommand{\rv}{{\mathbf{r}}}

\newcommand{\bing}[1]{\textcolor{black}{#1}}

\title{Nonresonant Casimir-Polder repulsion with a monolayer topological insulator}
\shorttitle{Nonresonant Casimir-Polder repulsion with a topological insulator} 

\author{Bing-Sui Lu\inst{1}}
\shortauthor{B.-S. Lu}

\institute{                    
  \inst{1} Departamento de Qu\'{i}mica, Universidad T\'{e}cnica Federico Santa Mar\'{i}a,
Campus San Joaquin, 7820275, Santiago, Chile
}

\abstract{
We investigate the behavior of the nonresonant Casimir-Polder force acting on a metastable polarized state of a two-level atomic system \bing{with a right-circularly polarized electric dipole transition} in the presence of a monolayer topological insulator inhabiting an anomalous quantum Hall insulator state with a negative Chern number, finding that the force can be repulsive for a certain range within the far-field region. By considering stanene irradiated by circularly polarized monochromatic laser light as a material example of a monolayer topological insulator, we study the effect of the frequency dispersion in the conductivity tensor on the force behavior, finding that the dispersion generally leads to a slight downward shift of the force relative to that predicted by a nondispersive approximation, as well as a slight reduction of the range over which the force is repulsive.}

\begin{document}

\maketitle

\section{Introduction}

In micro- and nano- electromechanical systems, the stiction of device components is a common issue and has been attributed to Casimir/van der Waals forces~\cite{allen2005,bordag}. Besides the stiction of mesoscopic material surfaces, atoms can also stick to surfaces owing to attractive Casimir-Polder forces. This can cause issues, for example, in the transport of atoms along hollow-core optical fibers~\cite{renn1995,afanasiev2010}, and leads to the loss of atoms in waveguides for matter waves~\cite{vorrath2010}. 
A possible way to reduce stiction is to make use of materials that exhibit repulsive Casimir forces. Time reversal symmetry-broken topological insulators are predicted to be an example of such materials~\cite{woods2016,khusnutdinov2019,lu2021,grushin2011a,grushin2011b,chen2011,chen2012,pablo2014,nie2013,zeng2016,pablo2017,fuchs2017a,lu2018,silveirinha2018,gangaraj2018}. The Casimir-Polder force can also turn repulsive under certain circumstances. For example, special types of the boundary surface geometry (such as a wedge apex, a screen with an aperture, or an annular dielectric) can lead to a repulsive nonresonant Casimir-Polder force if the atom is also anisotropically polarizable~\cite{milton2011,marchetta2021}. Furthermore, if an excited atom is near a planar surface, it experiences a resonant Casimir-Polder force which can oscillate with the atom-surface separation distance if the surface is made of a topological material, leading to repulsion for certain separation ranges~\cite{fuchs2017b,lu2022}. As the lifetime of an excited atomic state is typically of the order of nanoseconds, resonant Casimir-Polder repulsion may not be the best solution to reduce atom-surface stiction. 

\

For an atom near a planar surface, the nonresonant Casimir-Polder force is typically attractive if the atom is in a time reversal symmetry-unbroken ground state~\cite{casimir-polder1948,wylie-sipe1985}. 
On the other hand, if the atom is prepared in a metastable state that breaks time reversal symmetry, it can also experience a repulsive nonresonant Casimir-Polder force~\cite{lu2025}. In Ref.~\cite{lu2025}, we predicted that the nonresonant Casimir-Polder force acting on a polarized metastable lower-energy state of a two-level atomic system \bing{with a right-circularly polarized electric dipole transition}  can be repulsive for a certain range of separations if it is near a Chern insulator with a negative Chern number $C$. 
A Chern insulator is essentially a two-dimensional material tuned into the quantum anomalous Hall phase, for which the zero-frequency Hall conductance is integer quantized in units of $e^2/h$ (where $e$ is the electric charge, and $h$ is the Planck constant)~\cite{zhang2016}. Material candidates for Chern insulators include topological insulator thin films (such as $\rm{Bi_2Se_3}$) doped with transition metal atoms~\cite{chang2011,chang2013}, Floquet topological insulators such as monolayer graphene irradiated with circularly polarized monochromatic laser light~\cite{oka2009,kitagawa2011,cayssol2013}, and monolayer topological materials such as silicene, germanene and stanene, which have significant spin-orbit interaction strengths~\cite{ezawa2015}.  
Our prediction in Ref.~\cite{lu2025} was made based on the assumption that the conductivity tensor of the Chern insulator can be approximated by its zero frequency ({\it i.e.}, nondispersive) limit. To quadratic order in the fine-structure constant $\alpha$ (where $\alpha \approx 1/137$), the leading-order terms in an expansion of the inverse separation distance $1/d$ are found to be~\cite{lu2025} 
\be
f_{\rm{vdW}} \approx - \frac{\mu^2c}{\pi} \left( \frac{C^2\alpha^2}{ \omega_{10} d^5} + \frac{5c \, C\alpha}{\omega_{10}^2 d^6} \right), 
\label{fnd}
\ee
where $c$ is the speed of light, and $\omega_{10}$ is the electric dipole transition frequency of the two-level atomic system. 
The equation leads to the prediction that the force is repulsive if $C < 0$ and $d < 5c/(4|C|\alpha\omega_{10})$, and attractive if $C > 0$. 

\

One of the main objectives of the present paper is to determine whether the afore-mentioned prediction is still valid for a real Chern insulator material, for which the conductivity tensor  depends on frequency. 
For a realistic Chern insulator material system, we consider stanene which is irradiated by a circularly polarized monochromatic laser light, as stanene has a large spin-orbit interaction strength and the irradiation can also serve to enlarge the bandgap, which is helpful to achieving Casimir-Polder repulsion.  
 The plan of our paper is as follows. 
We start by providing an alternative derivation of the Casimir-Polder energy of a circularly polarized atomic state near a Chern insulator, using Lifshitz rarefaction. 
\bing{Next, we account for the frequency dependence of a Chern insulator material system by considering the low-energy description of stanene in the dissipationless limit. 
The use of the low-energy approximation also restricts our consideration of the force behavior to the far-field region. 
For the atom, we assume that it can be represented as a two-level system, with the lower energy state being a time reversal symmetry-broken metastable state that is connected to the higher energy state via an electric dipole transition. The metastable state typically has a lifetime of the order of seconds, which implies that on experimental timescales smaller than this, the metastable state behaves effectively like a ground state. 
Lastly, 
we study the behavior of the nonresonant Casimir-Polder force arising from different material parameter values of the Chern insulator.} 

\section{Diluting a Chern insulator}
\label{sec:dilute}

\begin{figure}[h]
\centering
  \includegraphics[width=0.47\textwidth]{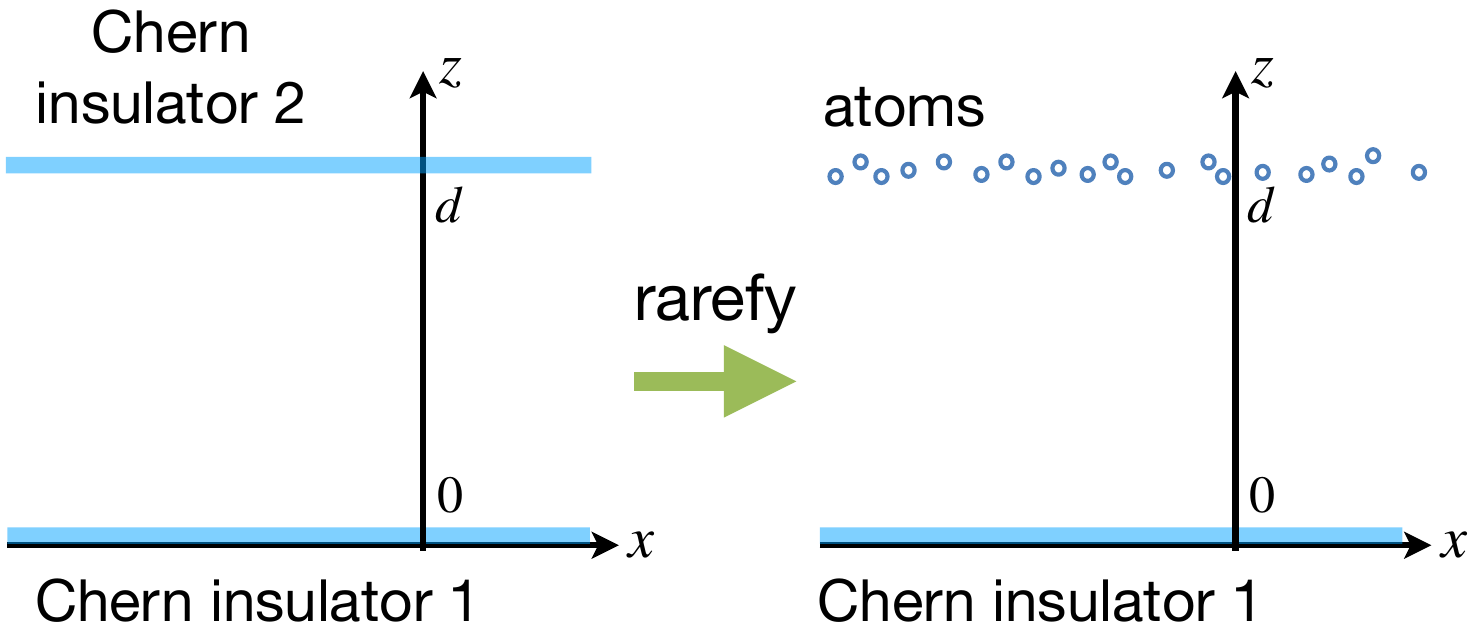}
  \caption{LEFT: Two Chern insulators separated by a distance $d$. RIGHT: To obtain the nonresonant Casimir-Polder energy, we dilute Chern insulator 2 into a gas of free atoms.} 
  \label{fig:CI}
\end{figure}

To obtain the nonresonant Casimir-Polder interaction energy between an atom and a  dielectric slab, one can either calculate (via quantum mechanical perturbation theory) the energy level shift in an atom induced by the presence of the dielectric~\cite{wylie-sipe1985}, or employ the method of ``Lifshitz rarefaction", in which one typically starts from the Casimir-Lifshitz (CL) energy between two dielectric slabs and dilute one of the slabs into a gas of free atoms~\cite{lifshitz1955,bordag}. The first method was used in Ref.~\cite{lu2025} to obtain the nonresonant Casimir-Polder energy of a circularly polarized atomic state near a Chern insulator, leading to Eq.~(9) in the same reference. Here, we shall show that Lifshitz rarefaction also leads to the same result. The zero-temperature CL energy for a pair of coplanar Chern insulators has already been obtained in Ref.~\cite{pablo2014}. Thus, rather than doing a fresh calculation of the CL energy between a Chern insulator and a dielectric slab, and diluting the slab into a gas of atoms, we work with the CL energy for two Chern insulators and dilute one of the Chern insulators into a gas of atoms. 
Since two Chern insulators with nonzero Hall conductivities can repel each other, the rarefaction derivation indicates the possibility that a circularly polarized atomic state and a Chern insulator can also repel each other.  

\

The idea behind Lifshitz rarefaction is to write the dielectric permittivity tensor $\bm{\varepsilon}$ in terms of the atomic polarizability $\bm{\alpha}$ of a gas of free atoms: 
\be
\label{eps1}
\bm{\varepsilon}(i\xi) \approx 
\mathbb{I} + \frac{4\pi N}{V} \bm{\alpha}(i\xi), 
\ee
where $\mathbb{I}$ is a $2\times2$ identity matrix, $N/V$ is the volumetric number density of atoms, and $i\xi$ is an imaginary frequency. 
For a Chern insulator, the optical response is given by its conductivity tensor $\bm{\sigma}$. 
In terms of $\bm{\varepsilon}$, we can write  
\be
\bm{\varepsilon}(i\xi) = \mathbb{I} + \frac{4\pi}{\xi} \delta(z) \bm{\sigma}(i\xi). 
\label{eps2}
\ee
Comparing Eq.~(\ref{eps1}) with Eq.~(\ref{eps2}), we see that Lifshitz rarefaction performed on a Chern insulator amounts to the following replacement:
\be
\bm{\sigma}(i\xi) \approx \frac{N}{A} \xi \bm{\alpha}(i\xi), 
\ee
 where $A$ is the area of the Chern insulator. 
 As $\sigma_{xx} = \sigma_{yy}$ and $\sigma_{xy} = - \sigma_{yx}$, we can decompose $\bm{\alpha}$ into diagonal and off-diagonal parts, {\it viz.}, $\bm{\alpha} = \bm{\alpha}^S + \bm{\alpha}^A$, where the off-diagonal part $\bm{\alpha}^A$ is antisymmetric: 
 \be
\bm{\alpha}^S = 
\begin{pmatrix}
\alpha_{xx} & 0 \\
0 & \alpha_{xx}
\end{pmatrix}, \,\,
\bm{\alpha}^A = 
\begin{pmatrix}
0 & \alpha_{xy} \\
- \alpha_{xy} & 0 
\end{pmatrix}, 
\ee
and we can write 
 \be
\sigma_{xx}(i\xi) = \frac{N}{A} \xi \alpha_{xx}(i\xi), \quad 
\sigma_{xy}(i\xi) = \frac{N}{A} \xi \alpha_{xy}(i\xi). 
 \ee
 The zero-temperature Casimir-Lifshitz energy for two coplanar Chern insulators separated by a distance $d$ in the $z$ direction (cf. Figure~\ref{fig:CI}) is given by 
\be
\label{CL}
\frac{E(d)}{A\hbar} = \int_0^\infty \!\frac{d\xi}{2\pi} \!\int_0^\infty \!\frac{dk_\perp k_\perp}{2\pi} \ln \det \Big( \mathbb{I} - \mathbb{R}(i\xi) \cdot \mathbb{R}'(i\xi) \, e^{-2 q d} \Big),  
\ee
where $\mathbb{I}$ is the identity matrix, $q = ((\xi/c)^2 + k_\perp^2)^{1/2}$, $k_\perp = (k_x^2 + k_y^2)^{1/2}$ is the magnitude of the transverse wavevector, $\mathbb{R}$ is the reflectivity matrix of the Chern insulator positioned at $z=0$, and $\mathbb{R}'$ is the reflectivity matrix of the Chern insulator positioned at $z=d$:  
\be
\mathbb{R} = 
\begin{pmatrix}
r_{ss} & r_{sp} 
\\
r_{ps} & r_{pp}
\end{pmatrix}, 
\quad 
\mathbb{R}' = 
\begin{pmatrix}
r_{ss}' & r_{sp}'
\\
r_{ps}' & r_{pp}'
\end{pmatrix}. 
\ee
The reflection coefficients for the first Chern insulator (with longitudinal and Hall conductivities $\sigma_{xx}$ and $\sigma_{xy}$) are given by 
\ba
r_{ss}(i\xi) &\!\!=\!\!& 
-
\frac{
\tsigma_{xx}^2 + \tsigma_{xy}^2
+ \frac{\xi}{c q} \tsigma_{xx} 
}{\tsigma_{xx}^2 + \tsigma_{xy}^2 + \big( \frac{\xi}{c q} + \frac{c q}{\xi} \big) \tsigma_{xx} + 1},
\nonumber\\
r_{ps}(i\xi) &\!\!=\!\!& 
r_{sp}(i\xi) 
= - \frac{\tsigma_{xy}}{\tsigma_{xx}^2 + \tsigma_{xy}^2 + \big( \frac{\xi}{c q} + \frac{c q}{\xi} \big) \tsigma_{xx} + 1}, 
\nonumber\\
r_{pp}(i\xi) &\!\!=\!\!& 
\frac{\tsigma_{xx}^2 + \tsigma_{xy}^2 
+ \frac{c q}{\xi} \tsigma_{xx}}{\tsigma_{xx}^2 + \tsigma_{xy}^2 + \big( \frac{\xi}{c q} + \frac{c q}{\xi} \big) \tsigma_{xx} + 1}, 
\ea
where we introduce the dimensionless longitudinal and Hall conductivities 
\be
\tsigma_{xx} \equiv (2\pi/c) \sigma_{xx}, \,\, \tsigma_{xy} \equiv (2\pi/c) \sigma_{xy}. 
\ee
The reflection coefficients for the second Chern insulator (with longitudinal and Hall conductances $\sigma_{xx}'$ and $\sigma_{xy}'$) are similar, except for a sign change in $r_{ps}'$ and $r_{sp}'$~\cite{lu2021}:  
\ba
r_{ss}'(i\xi) &\!\!=\!\!& 
-
\frac{
\tsigma_{xx}^{\prime 2} + \tsigma_{xy}^{\prime 2}
+ \frac{\xi}{c q} \tsigma_{xx}'
}{\tsigma_{xx}^{\prime 2} + \tsigma_{xy}^{\prime 2} + \big( \frac{\xi}{c q} + \frac{c q}{\xi} \big) \tsigma_{xx}' + 1},
\nonumber\\
r_{ps}'(i\xi) &\!\!=\!\!& 
r_{sp}'(i\xi) 
= \frac{\tsigma_{xy}'}{\tsigma_{xx}^{\prime 2} + \tsigma_{xy}^{\prime 2} + \big( \frac{\xi}{c q} + \frac{c q}{\xi} \big) \tsigma_{xx}' + 1}, 
\nonumber\\
r_{pp}'(i\xi) &\!\!=\!\!& 
\frac{\tsigma_{xx}^{\prime 2} + \tsigma_{xy}^{\prime 2} 
+ \frac{c q}{\xi} \tsigma_{xx}'}{\tsigma_{xx}^{\prime 2} + \tsigma_{xy}^{\prime 2} + \big( \frac{\xi}{c q} + \frac{c q}{\xi} \big) \tsigma_{xx}' + 1}.    
\ea
On diluting the second Chern insulator, the reflection coefficients are given to leading order in $N/A$ by 
\begin{subequations}
\ba
r_{ss}'(i\xi) &\!\!\approx\!\!& - \frac{2\pi\xi^2}{c^2q} \frac{N}{A} \alpha_{xx}(i\xi),  
\\
r_{ps}'(i\xi) &\!\!=\!\!& r_{sp}'(i\xi) \approx \frac{2\pi\xi}{c} \frac{N}{A} \alpha_{xy}(i\xi), 
\\
r_{pp}'(i\xi) &\!\!\approx\!\!& 2\pi q \frac{N}{A} \alpha_{xx}(i\xi).  
\ea
\end{subequations}
On substituting the above values into Eq.~(\ref{CL}), we find to linear order in $N/A$ that 
\ba 
\frac{E(d)}{A\hbar} 
&\!\!\approx\!\!& 
- \frac{2\pi N}{A}
\int_0^\infty \!\frac{d\xi}{2\pi} \int_0^\infty \! \frac{dk_\perp k_\perp}{2\pi}
\nonumber\\
&&\times 
\bigg( 
\left( - \frac{\xi^2}{c^2q} r_{ss}(i\xi) + q r_{pp}(i\xi) \right) \alpha_{xx}(i\xi) 
\nonumber\\
&&\quad 
+ \frac{\xi}{c} \big( r_{ps}(i\xi) + r_{sp}(i\xi) \big) \alpha_{xy}(i\xi) 
\bigg) \, 
e^{-2 q d}.  
\label{CL2}
\ea 
The above energy can be expressed in terms of the xx and xy components of the reflection Green tensor $G_{\alpha\beta}^R(\rv, \rv_0; \omega)$ (where $\rv_0 = (0,0,d)^{\rm T}$ is the position vector of the atom and $\rv$ is the field point), which are given for the case of imaginary frequencies by~\cite{lu2020,lu2022}
\ba
&&G^R_{xx}({\bf r}_0, {\bf r}_0, i\xi) 
\nonumber\\
&=&
-\frac{1}{2} \frac{\xi^2}{c^2} \! 
\int_{0}^\infty \!\!\!\! dk_\perp \frac{k_\perp}{q} \, e^{-2 q d} 
\big( 
r_{ss}(i\xi) - \frac{c^2 q^2}{\xi^2} r_{pp}(i\xi)
\big), 
\nonumber\\
&&G^R_{xy}({\bf r}_0, {\bf r}_0, i\xi) 
\nonumber\\
&=&
-\frac{1}{2} \Big( \frac{\xi}{c} \Big) \! 
\int_{0}^\infty \!\!\!\! dk_\perp k_\perp e^{- 2 q d} 
\left( r_{ps}(i\xi) + r_{sp}(i\xi) \right). 
\label{GR0}
\ea
For a single atom, $N=1$, and Eq.~(\ref{CL2}) then becomes 
\ba
E(d)
&\!\!=\!\!&
- \frac{\hbar}{\pi}
\int_0^\infty \!\!\! d\xi 
\bigg( 
\alpha_{xx}(i\xi) G_{xx}^R(\rv_0,\rv_0; i\xi) 
\nonumber\\
&&- \alpha_{xy}(i\xi) G_{xy}^R(\rv_0,\rv_0; i\xi) 
\bigg). 
\ea
Next, we consider a two-level atomic system with the lower-energy state denoted by $|0\rangle$ and the higher-energy state denoted by $|1\rangle$, and let us represent the electric dipole transition operator by $\bm{\mu}$.  
The polarizability tensor of state $|0\rangle$ is given by $\alpha_{\alpha\beta}^0 = \alpha_{\alpha\beta}^S + \alpha_{\alpha\beta}^A$, where $\alpha_{\alpha\beta}^S$ ($\alpha_{\alpha\beta}^A$) is the symmetric (antisymmetric) part~\cite{wylie-sipe1985}: 
\begin{subequations}
\label{alphaSA}
\ba
\alpha_{\alpha\beta}^S (\omega)
&\!\!=\!\!& 
\frac{1}{\hbar}
\frac{\omega_{10} ( \langle 0 |\mu_\alpha | 1 \rangle \langle 1 | \mu_\beta | 0 \rangle + \langle 0 |\mu_\beta | 1 \rangle \langle 1 | \mu_\alpha | 0 \rangle )}{\omega_{10}^2-\omega^2}, 
\nonumber\\
\alpha_{\alpha\beta}^A (\omega)
&\!\!=\!\!& 
\frac{1}{\hbar}
\frac{\omega ( \langle 0 |\mu_\alpha | 1 \rangle \langle 1 | \mu_\beta | 0 \rangle - \langle 0 |\mu_\beta | 1 \rangle \langle 1 | \mu_\alpha | 0 \rangle )}{\omega_{10}^2-\omega^2}. 
\nonumber
\ea
\end{subequations}
For our purpose, we assume that the state $|0\rangle$ is metastable and polarized to a nonzero angular momentum number ({\em e.g.}, via optical pumping). 
Let us consider a \bing{right-}circularly polarized \bing{electric} dipole transition described by $\langle 1 | \mu_x | 0 \rangle = -\mu/\sqrt{2}$ and $\langle 1 | \mu_y | 0 \rangle = - i\mu/\sqrt{2}$ (which  implies $\langle 0 | \mu_x | 1 \rangle = -\mu/\sqrt{2}$, $\langle 0 | \mu_y | 1 \rangle = i\mu/\sqrt{2}$), with the state $|1\rangle$ having a $m_J$ value larger than that of the state $| 0 \rangle$ by unity. 
We obtain $\alpha_{xx} = \omega_{10}\mu^2/(\omega_{10}^2 - \omega^2)$ and $\alpha_{xy} = i \omega\mu^2/(\omega_{10}^2 - \omega^2)$. 
With these values, the energy becomes 
\ba
&&\delta E_0^{\rm{vdw}}(d) 
\label{Evdw_0}
\\
&=& - \frac{\mu^2}{\pi} \int_0^\infty \!\!\!\!d\xi \Big( \frac{\omega_{10} \, G_{xx}^R(\rv_0,\rv_0; i\xi)}{\omega_{10}^2 + \xi^2} 
+ \frac{\xi \, G_{xy}^R(\rv_0,\rv_0; i\xi)}{\omega_{10}^2 + \xi^2} \Big), 
\nonumber
\ea
where we have now written $\delta E_0^{\rm{vdw}}$ instead of $E(d)$ for the nonresonant Casimir-Polder energy. We have recovered Eq.~(9) of Ref.~\cite{lu2025}, which was obtained by calculating the energy level shift of a circularly polarized atomic state induced by a Chern insulator.

\section{Conductivity tensor of monolayer stanene} 
\label{sec:conductivity}

To incorporate the effect of frequency dispersion and study how this modifies the nonresonant Casimir-Polder force behavior, we consider stanene, which has a buckled honeycomb monolayer structure with two sublattices, and the spin-orbit interaction opens up bandgaps at the $K_\eta$ points (with the valley index $\eta = +1$ corresponding to the $K$ point and $\eta = -1$ corresponding to the $K'$ point) for each spin~\cite{ezawa2015}. 
Via the photovoltaic Hall effect, the bandgaps can be tuned by shining a beam of monochromatic laser light on the monolayer~\cite{oka2009}. 
In the low-energy approximation, the dispersion of the electronic band structure near the Fermi level at the $K_\eta$ points for each spin polarization $s$ (where $s = +1$ corresponds to an up-spin and $s = -1$ corresponds to a down-spin) is described by a Dirac-like Hamiltonian. The bandgap is $2|\Delta_s^\eta|$, with $\Delta_s^\eta$ being the Dirac mass~\cite{ezawa2015}:
\be
\Delta_s^\eta = \eta s \lambda_{\rm{SO}} + \eta \lambda_H. 
\label{Deltaseta}
\ee
In the above equation, $\lambda_{\rm{SO}}$ is the spin-orbit interaction strength, and $\lambda_H$ is a mass-like term generated by the photovoltaic effect~\cite{kitagawa2011}. 
If we take the laser light to be described by a time-dependent vector potential ${\bf{A}}(t) = - A_0(\sin (\Omega t), \sin (\Omega t - \varphi))$, where 
$\Omega$ is the laser frequency, $\varphi$ is a phase angle, 
$v_F$ is the Fermi velocity, and $A_0$ is the amplitude, 
then $\lambda_H$ is given by~\cite{cayssol2013}
\be
\lambda_H = - \frac{(eA_0v_F)^2}{\hbar\Omega}\sin\varphi. 
\label{lambdaH}
\ee
In what follows, we set $\varphi = \pi/2$, which corresponds to left-circularly polarized laser light. 
Equation~(\ref{lambdaH}) is obtained in an approximation that requires $eA_0v_F/(\hbar\Omega)$ to be smaller than unity~\cite{cayssol2013}. 

\

Considering the case of zero dissipation and assuming that the Fermi level lies within the mass gap, the conductivities $\sigma_{xx}^{\eta,s}$ and $\sigma_{xy}^{\eta,s}$ for a Dirac cone with Dirac mass $\Delta_s^\eta$ are given for the case of imaginary frequencies by~\cite{pablo2017}
\ba
\sigma_{xx}^{\eta,s}(i\xi) &=& 
\frac{\alpha c}{4 \pi} \frac{|\Delta_s^\eta|}{\hbar \xi} 
+ 
\frac{\alpha c}{8 \pi} 
\bigg(
1 - \frac{4 (\Delta_s^\eta)^2}{\hbar^2 \xi^2} 
\bigg)
\tan^{-1} \bigg( \frac{\hbar \xi}{2 |\Delta_s^\eta|} \bigg), 
\nonumber\\
\sigma_{xy}^{\eta,s}(i\xi) &=& 
\frac{\alpha c}{2 \pi} \frac{\eta \Delta_s^\eta}{\hbar \xi} 
\tan^{-1} \bigg( \frac{\hbar \xi}{2 |\Delta_s^\eta|} \bigg).
\label{conduct}
\ea
In the above equation, $\alpha \equiv e^2/(\hbar c) \approx 1/137$ is the fine-structure constant, not to be confused with the atomic polarizability tensor $\bm{\alpha}$ which is represented in boldface. 
The total conductivity tensor is obtained by summing over the contributions from all Dirac cones: 
\be
\bm{\sigma}^{\rm{tot}} = \bm{\sigma}^{1,1}+\bm{\sigma}^{1,-1}+\bm{\sigma}^{-1,1}+\bm{\sigma}^{-1,-1}.
\label{sigma-tot}
\ee
To simplify the notation, we introduce the dimensionless frequency $X$, the dimensionless momentum $Y$, and the dimensionless conductivity tensor $\bm{\sigma}$: 
\ba
X &=& \frac{\hbar \xi}{m}, 
\,\, 
Y = \frac{\hbar c k_\perp}{m}, 
\,\,
\bm{\tsigma}^{\rm{tot}} = \frac{2\pi}{c}\bm{\sigma}^{\rm{tot}}, 
\nonumber\\
\frac{1}{m} &\equiv& 
\bigg( 
\frac{1}{|\Delta_1^1|} + \frac{1}{|\Delta_1^{-1}|} 
+ \frac{1}{|\Delta_{-1}^1|} + \frac{1}{|\Delta_{-1}^{-1}|}
\bigg).
\label{m-def}
\ea
In what follows, we look at the case of stanene, and obtain formulas for the longitudinal and Hall conductivities for the following values of $\lambda_H$: (i)~$\lambda_H = -\frac{3}{2} \lambda_{SO}$, (ii)~$\lambda_H = -\frac{5}{2} \lambda_{SO}$, and (iii)~$\lambda_H = -\frac{7}{2} \lambda_{SO}$. We have chosen these values, because they give rise to a quantum anomalous Hall phase with $C = - 2$. 

\subsection{$\lambda_H = -\frac{3}{2} \lambda_{SO}$}

First, we consider a circularly polarized laser shining on stanene, with phase angle $\varphi = \pi/2$, laser frequency $\Omega = 1000$ THz and 
$
\lambda_H = -\frac{3}{2} \lambda_{SO}. 
$
For stanene, the lattice constant is $a = 4.70 \rm{\AA}$, the Fermi velocity is $v_F = 4.9\times 10^5 \, {\rm m/s}$, and the spin-orbit strength is $\lambda_{SO} = 100\, {\rm meV}$~\cite{ezawa2015}. 
Using Eq.~(\ref{lambdaH}), we have $eA_0v_F/(\hbar\Omega) = \sqrt{|\lambda_H|/(\hbar\Omega)} \approx 0.48$. 
Our chosen value of $\lambda_H$ leads to 
$
\Delta_1^1 = -\frac{1}{2} \lambda_{SO}$, $\Delta_{1}^{-1} = \frac{1}{2} \lambda_{SO}$,
$\Delta_{-1}^{1} = -\frac{5}{2} \lambda_{SO}$, $\Delta_{-1}^{-1} = \frac{5}{2} \lambda_{SO}.
$ 
From Eq.~(\ref{m-def}), we obtain $m = 5\lambda_{SO}/24$. 
Using these values with Eqs.~(\ref{sigma-tot}), we find the total conductivity in terms of the dimensionless frequency $X$: 
\ba
\tsigma_{xx}^{\rm{tot}}(X) 
&\!\!=\!\!& 
\alpha \bigg(
\frac{72}{5X} 
+ \frac{1}{2} \tan^{-1} \Big( \frac{5X}{24} \Big)
+ \frac{1}{2} \tan^{-1} \Big( \frac{X}{24} \Big)
\nonumber\\
&&- \frac{288}{25X^2} \tan^{-1} \Big( \frac{5X}{24} \Big)
- \frac{288}{X^2} \tan^{-1} \Big( \frac{X}{24} \Big)
\bigg), 
\nonumber\\
\tsigma_{xy}^{\rm{tot}}(X) 
&\!\!=\!\!&
- \frac{\alpha}{X} \bigg(
\frac{24}{5} \tan^{-1} \Big( \frac{5X}{24} \Big) 
+ 
24 \tan^{-1} \Big( \frac{X}{24} \Big) 
\bigg).
\nonumber\\
\label{cond32}
\ea

\subsection{$\lambda_H = -\frac{5}{2} \lambda_{SO}$}

Next, we consider the same values for the laser frequency $\Omega$ and phase angle $\varphi$ as in the previous case, but with a different amplitude such that 
$
\lambda_H = -\frac{5}{2} \lambda_{SO}. 
$
For stanene, we have $eA_0v_F/(\hbar\Omega) = \sqrt{|\lambda_H|/(\hbar\Omega)} \approx 0.62$. 
Our value for $\lambda_H$ yields
$
\Delta_1^1 = -\frac{3}{2} \lambda_{SO}$, $\Delta_{1}^{-1} = \frac{3}{2} \lambda_{SO}$,
$\Delta_{-1}^{1} =-\frac{7}{2} \lambda_{SO}$, $\Delta_{-1}^{-1} = \frac{7}{2} \lambda_{SO}$. 
From Eq.~(\ref{m-def}), we have $m = 21\lambda_{SO}/40$. 
Using these values together with Eqs.~(\ref{sigma-tot}), we obtain the total conductivity 
in terms of the dimensionless frequency $X$: 
\ba
\tsigma_{xx}^{\rm{tot}}(X) 
&\!\!=\!\!& 
\alpha \bigg(
\frac{200}{21X} 
+ \frac{1}{2} \tan^{-1} \Big( \frac{7X}{40} \Big)
+ \frac{1}{2} \tan^{-1} \Big( \frac{3X}{40} \Big)
\nonumber\\
&&- \frac{800}{49X^2} \tan^{-1} \Big( \frac{7X}{40} \Big)
- \frac{800}{9X^2} \tan^{-1} \Big( \frac{3X}{40} \Big)
\bigg), 
\nonumber\\
\tsigma_{xy}^{\rm{tot}}(X) 
&\!\!=\!\!&
- \frac{\alpha}{X} \bigg(
\frac{40}{7} \tan^{-1} \Big( \frac{7X}{40} \Big) 
+ 
\frac{40}{3} \tan^{-1} \Big( \frac{3X}{40} \Big) 
\bigg).
\nonumber\\
\label{cond52}
\ea

\subsection{$\lambda_H = -\frac{7}{2} \lambda_{SO}$}

Lastly, we again consider the same values for the laser frequency $\Omega$ and phase angle $\varphi$ as in the previous two cases, but with an amplitude such that
$
\lambda_H = -\frac{7}{2} \lambda_{SO}. 
$
For stanene, this gives $eA_0v_F/(\hbar\Omega) = \sqrt{|\lambda_H|/(\hbar\Omega)} \approx 0.73$. 
We have 
$
\Delta_1^1 = -\frac{5}{2} \lambda_{SO}$, $\Delta_{1}^{-1} = \frac{5}{2} \lambda_{SO}$,
$\Delta_{-1}^{1} =-\frac{9}{2} \lambda_{SO}$, $\Delta_{-1}^{-1} = \frac{9}{2} \lambda_{SO}$.
From Eq.~(\ref{m-def}), we have $m = 45\lambda_{SO}/56$.
Using these values together with Eqs.~(\ref{sigma-tot}), we obtain the total conductivity in terms of the dimensionless frequency $X$: 
\ba
\tsigma_{xx}^{\rm{tot}}(X) 
&\!\!=\!\!& 
\alpha \bigg(
\frac{392}{45X} 
+ \frac{1}{2} \tan^{-1} \Big( \frac{9X}{56} \Big)
+ \frac{1}{2} \tan^{-1} \Big( \frac{5X}{56} \Big)
\nonumber\\
&&- \frac{1568}{81X^2} \tan^{-1} \Big( \frac{9X}{56} \Big)
- \frac{1568}{25X^2} \tan^{-1} \Big( \frac{5X}{56} \Big)
\bigg), 
\nonumber\\
\tsigma_{xy}^{\rm{tot}}(X) 
&\!\!=\!\!&
- \frac{\alpha}{X} \bigg(
\frac{56}{9} \tan^{-1} \Big( \frac{9X}{56} \Big) 
+ 
\frac{56}{5} \tan^{-1} \Big( \frac{5X}{56} \Big) 
\bigg).
\nonumber\\
\label{cond72}
\ea
For each of the Eqs.~(\ref{cond32}), (\ref{cond52}) and (\ref{cond72}), we can check that to linear order in $X$, $\tsigma_{xx}^{\rm{tot}} \approx \alpha X/6$ and $\tsigma_{xy}^{\rm{tot}} \approx -2 \alpha$, consistent with the system residing in a $C = -2$ anomalous quantum Hall insulator state. 

\section{Nonresonant Casimir-Polder force behavior} 
\label{sec:force}

We now turn to study the nonresonant Casimir-Polder force behavior for the $C = - 2$ quantum anomalous Hall phase, for which the force acting on a metastable polarized atomic state can be repulsive. 
To study the nonresonant Casimir-Polder force, we need the reflection coefficients and reflection Green tensor. 
Expressed in terms of the dimensionless quantities, the reflection coefficients are given by~\cite{pablo2014,lu2020,lu2021,lu2022} 
\ba
&&r_{ss}(X) 
\nonumber\\
&\!\!=\!\!& 
-
\frac{
(\tsigma_{xx}^{\rm{tot}})^2 + (\tsigma_{xy}^{\rm{tot}})^2
+ \frac{X}{\sqrt{X^2+Y^2}} \tsigma_{xx}^{\rm{tot}} 
}{(\tsigma_{xx}^{\rm{tot}})^2 + (\tsigma_{xy}^{\rm{tot}})^2 + \big( \frac{X}{\sqrt{X^2+Y^2}} + \frac{\sqrt{X^2+Y^2}}{X} \big) \tsigma_{xx}^{\rm{tot}} + 1},
\nonumber\\
&&r_{ps}(X) 
= r_{sp}(X) 
\nonumber\\
&\!\!=\!\!& 
- \frac{\tsigma_{xy}^{\rm{tot}}}{(\tsigma_{xx}^{\rm{tot}})^2 + (\tsigma_{xy}^{\rm{tot}})^2 + \big( \frac{X}{\sqrt{X^2+Y^2}} + \frac{\sqrt{X^2+Y^2}}{X} \big) \tsigma_{xx}^{\rm{tot}} + 1}, 
\nonumber\\
&&r_{pp}(X) 
\nonumber\\
&\!\!=\!\!& 
\frac{(\tsigma_{xx}^{\rm{tot}})^2 + (\tsigma_{xy}^{\rm{tot}})^2 
+ \frac{\sqrt{X^2+Y^2}}{X} \tsigma_{xx}^{\rm{tot}}}{(\tsigma_{xx}^{\rm{tot}})^2 + (\tsigma_{xy}^{\rm{tot}})^2 + \big( \frac{X}{\sqrt{X^2+Y^2}} + \frac{\sqrt{X^2+Y^2}}{X} \big) \tsigma_{xx}^{\rm{tot}} + 1}. 
\nonumber\\
\label{refcoeffs}
\ea
We also define a dimensionless distance $D$, {\it viz.}, 
\be
D = 
\frac{m}{\hbar c} d,   
\ee
where $\hbar c/m$ provides a lengthscale for distinguishing the near- and far-field regions. 
In terms of dimensionless quantities, the Green tensor components in Eq.~(\ref{GR0}) are 
\ba
&&G^R_{xx}(D,X) 
=
-\frac{1}{2} \left( \frac{m}{\hbar c} \right)^3 X^2 
\int_{0}^\infty \!\!\!\! dY \, Y 
\nonumber\\
&&\quad\times
\frac{e^{-2 \sqrt{X^2+Y^2} D} }{\sqrt{X^2+Y^2}}   
\bigg( 
r_{ss}(X) - \left( 1 + \frac{Y^2}{X^2} \right) r_{pp}(X)
\bigg), 
\nonumber\\
&&G^R_{xy}(D,X)
=
-\frac{1}{2} \left( \frac{m}{\hbar c} \right)^3 X 
\int_{0}^\infty \!\!\!\! dY \, Y 
\nonumber\\
&&\quad\times
 e^{-2 \sqrt{X^2+Y^2} D} 
\left( r_{ps}(X) + r_{sp}(X) \right). 
\ea
If we define a dimensionless transition frequency 
\be
W \equiv \frac{\hbar\omega_{10}}{m}, 
\ee
we can re-express Eq.~(\ref{Evdw_0}) as
\be
\delta E_0^{{\rm vdw}}
=
- \frac{\mu^2}{\pi} 
\int_0^\infty \!\!\!\!\! \ dX
  \Bigg(
  \frac{W \, G_{xx}^R(D,X)}{W^2 + X^2}  
  + \frac{X \, G_{xy}^R(D,X)}{W^2 + X^2} 
  \Bigg).
\label{E0vdw-nondim}
\ee
\begin{figure}[h]
\centering
  \includegraphics[width=0.47\textwidth]{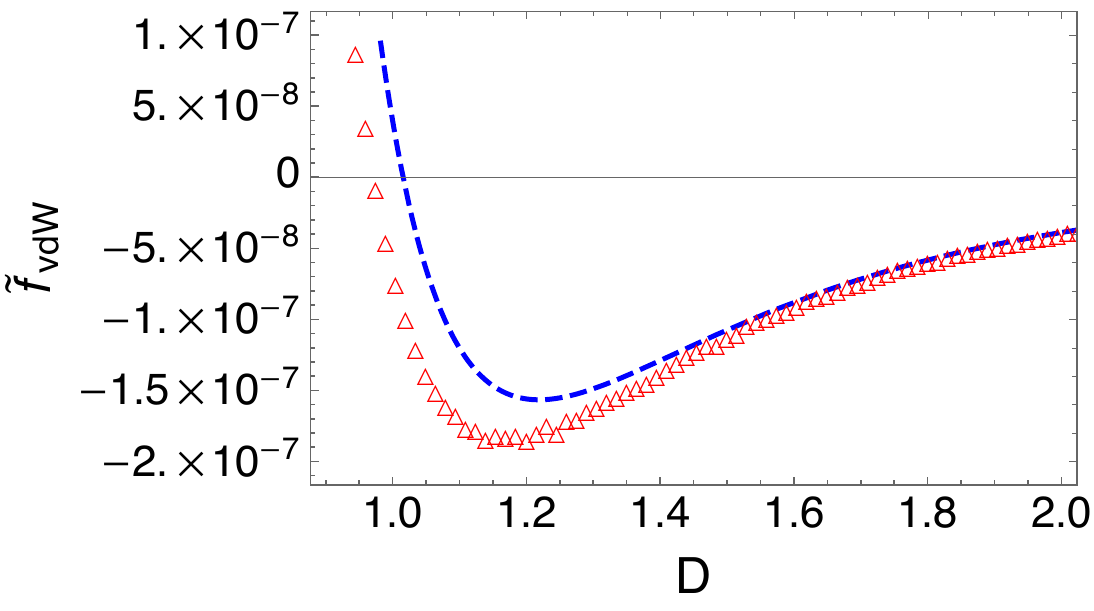}
  \caption{The dimensionless nonresonant Casimir-Polder force $\tilde{f}_{\rm{vdW}}$ as a function of the dimensionless separation $D$ for parameter values $\lambda_H = -3\lambda_{SO}/2, W = 84.3$, numerically plotted using the frequency dependent conductivity tensor Eq.~(\ref{conduct}) (red triangles) and using the nondispersive approximation Eq.~(\ref{force-nd}) (blue dashed line).} 
  \label{fig:plot32}
\end{figure}
\begin{figure}[h]
\centering
  \includegraphics[width=0.47\textwidth]{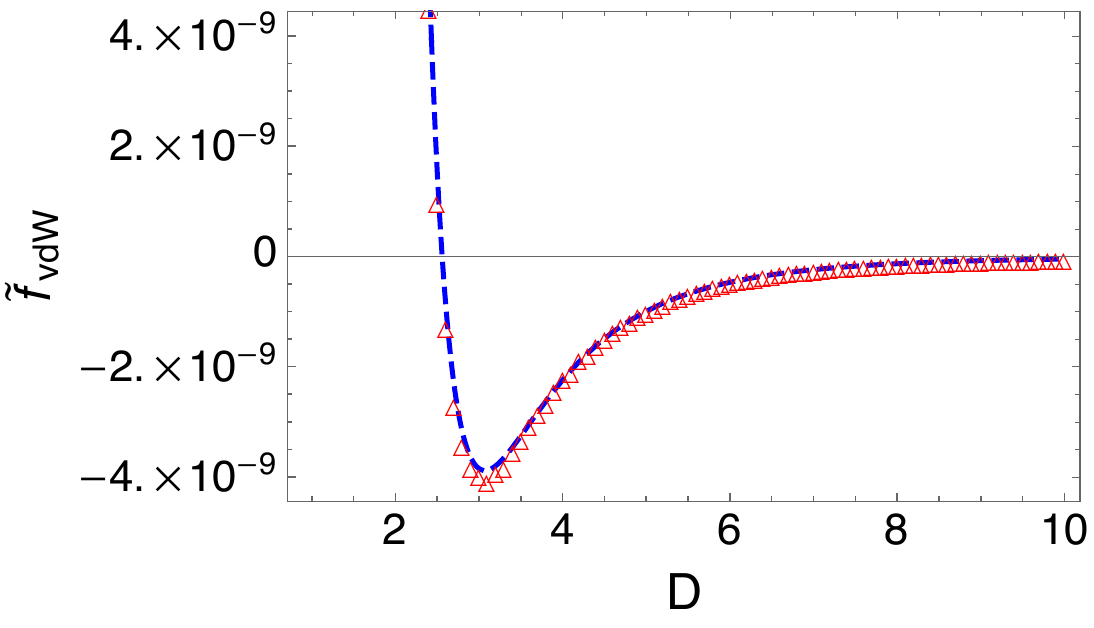}
  \caption{The dimensionless nonresonant Casimir-Polder force $\tilde{f}_{\rm{vdW}}$ as a function of the dimensionless separation $D$ for parameter values $\lambda_H = -5\lambda_{SO}/2, W = 33.45$, numerically plotted using the frequency dependent conductivity tensor Eq.~(\ref{conduct}) (red triangles) and using the nondispersive approximation Eq.~(\ref{force-nd}) (blue dashed line).} 
  \label{fig:plot52}
\end{figure}
\begin{figure}[h]
\centering
  \includegraphics[width=0.47\textwidth]{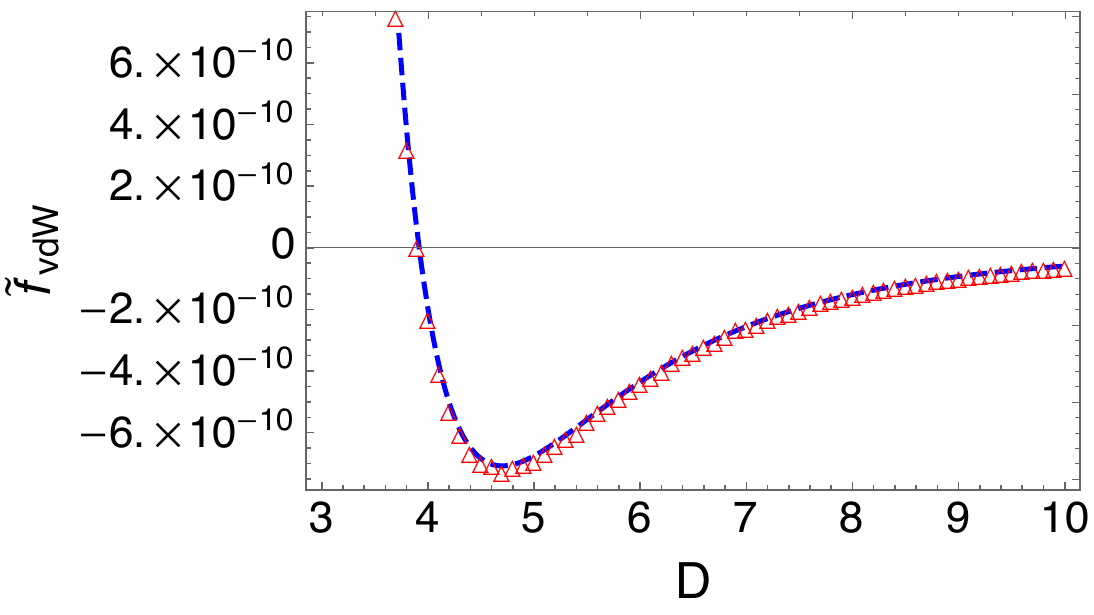}
  \caption{The dimensionless nonresonant Casimir-Polder force $\tilde{f}_{\rm{vdW}}$ as a function of the dimensionless separation $D$ for parameter values $\lambda_H = -7\lambda_{SO}/2, W = 21.86$, numerically plotted using the frequency dependent conductivity tensor Eq.~(\ref{conduct}) (red triangles) and using the nondispersive approximation Eq.~(\ref{force-nd}) (blue dashed line).} 
  \label{fig:plot72}
\end{figure}

The nonresonant Casimir-Polder force is given by 
\be
f_{\rm{vdW}} = - \frac{\partial \delta E_0^{{\rm vdw}}}{\partial d}. 
\ee
Defining the dimensionless force 
\be
\widetilde{f}_{\rm{vdW}} \equiv \frac{\pi \hbar^4 c^4}{\mu^2 m^4} f_{\rm{vdW}},  
\label{fnondim}
\ee
we obtain 
\ba
\widetilde{f}_{\rm{vdW}} 
&\!\!=\!\!& 
 \int_0^\infty \!\!\!\!\! dX
   \frac{W X^2}{W^2 + X^2}  
\int_{0}^\infty \!\!\!\!\! dY  \, Y \, e^{-2 \sqrt{X^2+Y^2} D} 
 \nonumber\\
&&\quad\times   
\bigg( 
r_{ss}(X) - \left( 1 + \frac{Y^2}{X^2} \right) r_{pp}(X)
\bigg)
\nonumber\\
&&+
 \int_0^\infty \!\!\!\!\! dX
   \frac{X^2}{W^2 + X^2}  
\int_{0}^\infty \!\!\!\! dY \, Y e^{-2 \sqrt{X^2+Y^2} D} 
 \nonumber\\
&&\quad\times   
\sqrt{X^2+Y^2} \left( r_{ps}(X) + r_{sp}(X) \right).  
\ea
In Figs.~\ref{fig:plot32}, \ref{fig:plot52} and \ref{fig:plot72}, we use the frequency dependent conductivity tensor Eq.~(\ref{conduct}) to numerically plot the distance behavior of the dimensionless nonresonant Casimir-Polder force $\tilde{f}_{\rm{vdW}}$ for (i)~$\lambda_H = - 3\lambda_{SO}/2, W = 84.3$, (ii)~$\lambda_H = - 5\lambda_{SO}/2, W=33.45$, and (iii)~$\lambda_H = - 7\lambda_{SO}/2, W=21.86$ respectively. 
The values of $W$ for these cases are obtained using the spin-orbit interaction strength of stanene ($\lambda_{SO} = 100$ meV) and the electric dipole transition wavelength ($707.202$ nm) of a two-level atomic system whose lower and higher energy states are respectively the $m_J = - 2$ polarized 5s5p $^3P_2$ state and the $m_J = - 1$ polarized 5s6s $^3S_1$ state of Strontium-88~\cite{schkolnik2019,xu2003}. 
In the figures, we also compare the force obtained using the frequency dependent conductivity with the force in Eq.~(\ref{fnd}), obtained from a nondispersive approximation. Using Eqs.~(\ref{fnd}) and (\ref{fnondim}), the dimensionless nonresonant Casimir-Polder force in the nondispersive approximation is given by 
\be
\tilde{f}_{\rm{vdW}} \approx - \frac{C^2 \alpha^2}{WD^5} - \frac{5C\alpha}{4W^2D^6}. 
\label{force-nd}
\ee
This equation tells us that independent of the sign of $C$, the force is always attractive at sufficiently large separations $D$, which agrees with what we see from Figs.~\ref{fig:plot32}, \ref{fig:plot52} and \ref{fig:plot72}. 
For $C < 0$, the force is repulsive, and Eq.~(\ref{force-nd}) allows an estimate of the distance $D_\ast$ at which the force changes sign. In the nondispersive approximation,   
\be
D_\ast = \frac{5}{4|C|\alpha W}. 
\label{dastnd}
\ee
The value of $D_\ast$ gives the size of the range of separations over which the nonresonant Casimir-Polder force is repulsive, the range being smaller for larger values of $W$ and/or $C$. 
As our use of the low-energy approximation constrains the validity of our force calculation to the far-field region, we also require for consistency that $D_\ast > 1$. 

\

From Figs.~\ref{fig:plot52} and \ref{fig:plot72}, we see that Eq.~(\ref{dastnd}) gives a good estimate of the distance at which the dispersive force changes sign. The dispersive force for $\lambda_{H} = -5\lambda_{SO}/2, W=33.45$ changes sign at $D = 2.54$, which is close to the value $D_\ast = 2.56$ predicted by the nondispersive approximation Eq.~(\ref{dastnd}). 
The dispersive force for $\lambda_{H} = -7\lambda_{SO}/2, W=21.86$ changes sign at $D = 3.90$, and this is also close to the value $D_\ast = 3.92$ predicted by the nondispersive approximation. 

\

On the other hand, as we see from Fig.~\ref{fig:plot32}, the nondispersive approximation predicts that the force vanishes at $D_\ast = 1.016$ (greater than 1), whereas the dispersive force vanishes at $D \approx 0.97$ (smaller than 1). In this case, the effect of frequency dispersion is to make the force attractive for the entirety of the far-field region. 
In each of the Figs.~\ref{fig:plot32}, \ref{fig:plot52} and \ref{fig:plot72}, we see that the dispersive force curve is always downward-shifted relative to the nondispersive force curve. 
The downward shift indicates that the effect of frequency dispersion is to enhance the attraction if the force is attractive, and reduce the repulsion if the force is repulsive. 
The downward shift also appears to be more pronounced for larger values of $W$.

\section{Summary and conclusion} 

In this paper, we derived the nonresonant Casimir-Polder energy of a two-level atomic system in a metastable, circularly polarized state near a Chern insulator by using Lifshitz rarefaction, which agrees with the formula obtained in Ref.~\cite{lu2025} using quantum mechanical perturbation theory. We also investigated the effect of frequency dispersion of the conductivity tensor on the nonresonant Casimir-Polder force behavior in the far-field region, for a two-level atomic system prepared in a metastable polarized state placed near stanene in a quantum anomalous Hall phase with Chern number $C = -2$ in the dissipationless limit. 
For the two-level atomic system, we have considered the \bing{right-circularly polarized} electric dipole transition between the $m_J = - 2$ polarized 5s5p $^3P_2$ state and the $m_J = - 1$ polarized 5s6s $^3S_1$ state of Strontium-88. 
By studying the force behavior for the three cases $\lambda_H = -3\lambda_{SO}/2$, $\lambda_H = -5\lambda_{SO}/2$, and $\lambda_H = -7\lambda_{SO}/2$ (where $\lambda_H$ is a photoinduced mass-like term, and $\lambda_{SO}$ is the spin-orbit interaction strength of stanene), we found the following results: 
\begin{itemize}
\item  The nonresonant Casimir-Polder force can become repulsive within a certain range in the far-field region. 
\item Frequency dispersion results in a downward shift of the nonresonant Casimir-Polder force relative to the nondispersive approximation. 
\item The downward shift is greater for larger values of the dimensionless electric dipole transition frequency $W = \hbar \omega_{10}(|\Delta_{1}^{1}|^{-1} + |\Delta_{-1}^{1}|^{-1} + |\Delta_{1}^{-1}|^{-1} + |\Delta_{-1}^{-1}|^{-1})$ (where $\Delta_s^\eta$ is the Dirac mass at the $K_\eta$ point for valley index $\eta$ and spin polarization $s$, and $\omega_{10}$ is the electric dipole transition frequency of the two-level atomic system). 
\item Frequency dispersion causes the range over which the force is repulsive to be slightly smaller than that predicted by a nondispersive approximation.  
\item The nondispersive approximation is valid for the cases $\lambda_H = -5\lambda_{SO}/2$ and $\lambda_H = -7\lambda_{SO}/2$, but not valid for $\lambda_H = -3\lambda_{SO}/2$, as the dispersive force behavior turns out to be attractive in the entirety of the far-field region. 
\end{itemize}
In conclusion, we have found that it is possible to engineer a repulsive nonresonant Casimir-Polder force in the far-field region on a metastable polarized state of a two-level atomic system \bing{with a right-circularly polarized electric dipole transition}, if it is within a certain separation range from a monolayer topological insulator inhabiting an anomalous quantum Hall state with a negative Chern number, and $\lambda_H$ is sufficiently negative and $W$ is sufficiently small. 
\bing{Our investigation can be relevant to the search for materials that reduce atom-surface stiction, and also suggests a potential method of determining the Chern number of a given topological material via a Casimir-Polder force measurement in the far-field region.}

\acknowledgments
The author thanks Galina Klimchitskaya and Vladimir Mostepanenko for the invitation to contribute to EPL's Focus Issue on ``Casimir Effect and Its Role in Modern Physics," and David Wilkowski for constructive discussions. In addition, he thanks Derek Frydel for having hosted him at the Universidad T\'{e}cnica Federico Santa Mar\'{i}a, where the present work was completed, and acknowledges financial support from FONDECYT through grant number 1241694, which made the visit possible.


\begin{thebibliography}{99}

\bibitem{allen2005} 
\Name{Allen J. J.} 
\Book{Micro Electro Mechanical System Design} 
\Publ{CRC Press, New York}
\Year{2005}.

\bibitem{bordag}
\Name{M. Bordag et al.}
\Book{Advances in the Casimir Effect (1st edition)}
\Publ{Oxford University Press, Oxford}
\Year{2009}. 

\bibitem{lifshitz1955} 
\Name{Lifshitz E. M.} 
\REVIEW{Sov. Phys. JETP}{2}{1956}{73}.

\bibitem{renn1995}
\Name{Renn M. J. et al.}
\REVIEW{Phys. Rev. Lett.}{75}{1995}{3253}. 

\bibitem{afanasiev2010}
\Name{Afanasiev A. \and Minogin V.} 
\REVIEW{Phys. Rev. A}{82}{2010}{052903}. 

\bibitem{vorrath2010}
\Name{Vorrath S. et al.} 
\REVIEW{New J. Phys.}{12}{2010}{123015}.   

\bibitem{woods2016} 
\Name{Woods L. M. et al}
\REVIEW{Rev. Mod. Phys.}{88}{2016}{045003}. 

\bibitem{khusnutdinov2019}
\Name{Khusnutdinov N. \and Woods L. M.}
\REVIEW{JETP Lett.}{110}{2019}{183}. 

\bibitem{lu2021}
\Name{Lu B.-S.}
\REVIEW{Universe}{7}{2021}{237}.

\bibitem{grushin2011a}
\Name{Grushin A. G. \and Cortijo A.}
\REVIEW{Phys. Rev. Lett.}{106}{2011}{020403}. 

\bibitem{grushin2011b}
\Name{Grushin A. G., Rodriguez-Lopez P. \and Cortijo A.}
\REVIEW{Phys. Rev. B}{84}{2011}{045119}.

\bibitem{chen2011} 
\Name{Chen L. \and Wan S.}  
\REVIEW{Phys. Rev. B}{84}{2011}{075149}. 

\bibitem{chen2012}  
\Name{Chen L. \and Wan S.} 
\REVIEW{Phys. Rev. B}{85}{2012}{115102}. 

\bibitem{nie2013}
\Name{Nie W et al.}
\REVIEW{Phys. Rev. B}{88}{2013}{085421}.

\bibitem{pablo2014}
\Name{Rodriguez-Lopez P. \and Grushin A. G.} 
\REVIEW{Phys. Rev. Lett.}{112}{2014}{056804}. 

\bibitem{zeng2016}
\Name{Zeng R. et al.}
\REVIEW{Phys. Lett. A}{380}{2016}{2861}. 

\bibitem{fuchs2017a}
\Name{Fuchs S. et al.}
\REVIEW{Phys. Rev. A}{96}{2017}{062505}.

\bibitem{pablo2017}
\Name{Rodriguez-Lopez P. et al.}
\REVIEW{Nat. Commun.}{8}{2017}{14699}. 

\bibitem{lu2018}
\Name{Lu B.-S.}
\REVIEW{Phys. Rev. B}{97}{2018}{045427}. 
\bing{
\bibitem{silveirinha2018}
\Name{Silveirinha M. G., Gangaraj S. A. H., Hanson G. W. \and Antezza M.}
\REVIEW{Phys. Rev. A}{97}{022509}. 
}
\bing{
\bibitem{gangaraj2018}
\Name{Gangaraj S. A. H., Hanson G. W., Antezza M. \and Silveirinha M. G.}
\REVIEW{Phys. Rev. B}{97}{201108(R)}.  
}

\bibitem{milton2011}
\Name{Milton K. et al.}
\REVIEW{Phys. Rev. A}{83}{2011}{062507}.

\bibitem{marchetta2021}
\Name{Marchetta J. J., Parashar P., \and Shajesh K. V.}
\REVIEW{Phys. Rev. A}{104}{2021}{032209}. 

\bibitem{fuchs2017b}
\Name{Fuchs S., Crosse J. A. \and Buhmann S. Y.}
\REVIEW{Phys. Rev. A}{95}{2017}{023805}. 

\bibitem{lu2022}
\Name{Lu B.-S., Arifa K. Z. \and Ducloy M.}
\REVIEW{Eur. Phys. J. D}{76}{2022}{210}. 

\bibitem{casimir-polder1948}
\Name{Casimir H. B. G. \and Polder D.}
\REVIEW{Phys. Rev.}{73}{1948}{360}. 

\bibitem{wylie-sipe1985}
\Name{Wylie J. M. \and Sipe J. E.}
\REVIEW{Phys. Rev. A}{32}{1985}{2030}.

\bibitem{lu2025} 
\Name{Lu B.-S.}
\REVIEW{Int. J. Mod. Phys. A}{40}{2025}{2543011}. 

\bibitem{zhang2016}
\Name{Zhang J. et al.}
\REVIEW{Chin. Phys. B}{25}{2016}{117308}. 

\bibitem{chang2011}
\Name{Chang C.-Z. et al.} 
\REVIEW{SPIN}{1}{2011}{21}. 

\bibitem{chang2013}
\Name{Chang C.-Z. et al.} 
\REVIEW{Science}{340}{2013}{167}. 

\bibitem{oka2009}
\Name{Oka T. \and Aoki H.} 
\REVIEW{Phys Rev. B}{79}{2009}{081406(R)}.

\bibitem{kitagawa2011}
\Name{Kitagawa T. et al.}
\REVIEW{Phys. Rev. B}{84}{2011}{235108}. 

\bibitem{cayssol2013} 
\Name{Cayssol J. et al.}
\REVIEW{Phys. Status Solidi RRL}{7}{2013}{101}.

\bibitem{ezawa2015} 
\Name{Ezawa M.} 
\REVIEW{J. Phys. Soc. Japan}{84}{2015}{121003}. 

\bibitem{lu2020} 
\Name{Lu B.-S., Arifa K. Z. \and Hong X. R.}
\REVIEW{Phys. Rev. B}{101}{2020}{205410}.

\bibitem{schkolnik2019}
\Name{Schkolnik V., Fartmann O. \and Krutzik M.}
\REVIEW{Laser Phys.}{29}{2019}{035802}.

\bibitem{xu2003}
\Name{Xu X. et al.}
\REVIEW{J. Opt. Soc. Am. B}{20}{2003}{968}.

\end{thebibliography}
\end{document}